\documentclass[10pt,letterpaper]{article}

\usepackage{cogsci}
\usepackage{pslatex}
\usepackage{apacite}
\usepackage{dblfloatfix}
\usepackage{subfig}

\usepackage{amsmath}
\usepackage{amsfonts}
\usepackage{amssymb}
\usepackage{graphicx}
\usepackage[round]{natbib}
\usepackage{color}

\newcommand{\lir}{\overset{l}{\rightsquigarrow}}
\newcommand{\cir}{\overset{c}{\rightsquigarrow}}

\newcommand{\bP}{\mathbb P}

\title{Bringing Order to the Cognitive Fallacy Zoo}

\author{{\large \bf Ardavan~S.~Nobandegani$^{1,3}$,  William~Campoli$^{2}$, \& Thomas~R.~Shultz$^{2,3}$}\\
\{ardavan.salehinobandegani, william.campoli\}@mail.mcgill.ca\\
\{thomas.shultz\}@mcgill.ca\vspace*{3pt}\\
\small{$^{1}$Department of Electrical \& Computer Engineering, McGill University}\\
\small{$^{2}$School of Computer Science, McGill University}\\
\small{$^{3}$Department of Psychology, McGill University}}

\begin{document}
\maketitle

\begin{abstract}
In the eyes of a rationalist like  Descartes or Spinoza, human reasoning is flawless, marching toward uncovering  ultimate truth. A few centuries later, however, culminating in the work of Kahneman and Tversky, human reasoning was portrayed as anything but flawless, filled with numerous misjudgments, biases, and cognitive fallacies. With further investigations, new cognitive fallacies continually emerged, leading to a state of affairs which can fairly be characterized as the cognitive fallacy zoo!~In this largely methodological work, we formally present a principled way to bring order to this zoo. We introduce the idea of establishing {implication relationships} (IRs) between cognitive fallacies, formally characterizing how one fallacy implies another. IR is analogous to, and partly inspired by, the fundamental concept of {reduction} in computational complexity theory. We present several examples of IRs involving experimentally well-documented cognitive fallacies: base-rate neglect, availability bias, conjunction fallacy, decoy effect, framing effect, and Allais paradox. We conclude by discussing how our work: (i) allows for identifying those pivotal cognitive fallacies whose investigation would be the most rewarding research agenda, and importantly (ii) permits a systematized, guided research program on cognitive fallacies, motivating influential theoretical as well as experimental avenues of future research.

\end{abstract}

\section{Introduction}
The Enlightenment was the golden age of human rationality. In the eyes of a rationalist like  Descartes or Spinoza, human reasoning is flawless, marching toward uncovering ultimate truth. Adopting the view that human experience is, at best, partial, and at worst misleading, human reasoning was held by rationalists to be \emph{the} means for attaining truth. In Kantian terminology, rationalists maintained that human {a posteriori} knowledge is prone to imperfections, but {a priori} knowledge is free of such impurities. Upon careful empirical investigations, however, human reasoning was found to systematically deviate from normative principles like the laws of probability and logic in numerous ways, thus strongly challenging rationalism. A few centuries since the birth of rationalism, human reasoning is now portrayed as anything but flawless, filled with numerous misjudgments, biases, and cognitive fallacies (e.g., Simon, 1972; Kahneman \& Tversky, 1974; Kahneman, 2011).

\begin{figure*}[t!]
\centering
\includegraphics[trim = 16mm 0mm 16mm 20mm, clip, width=.97\textwidth]{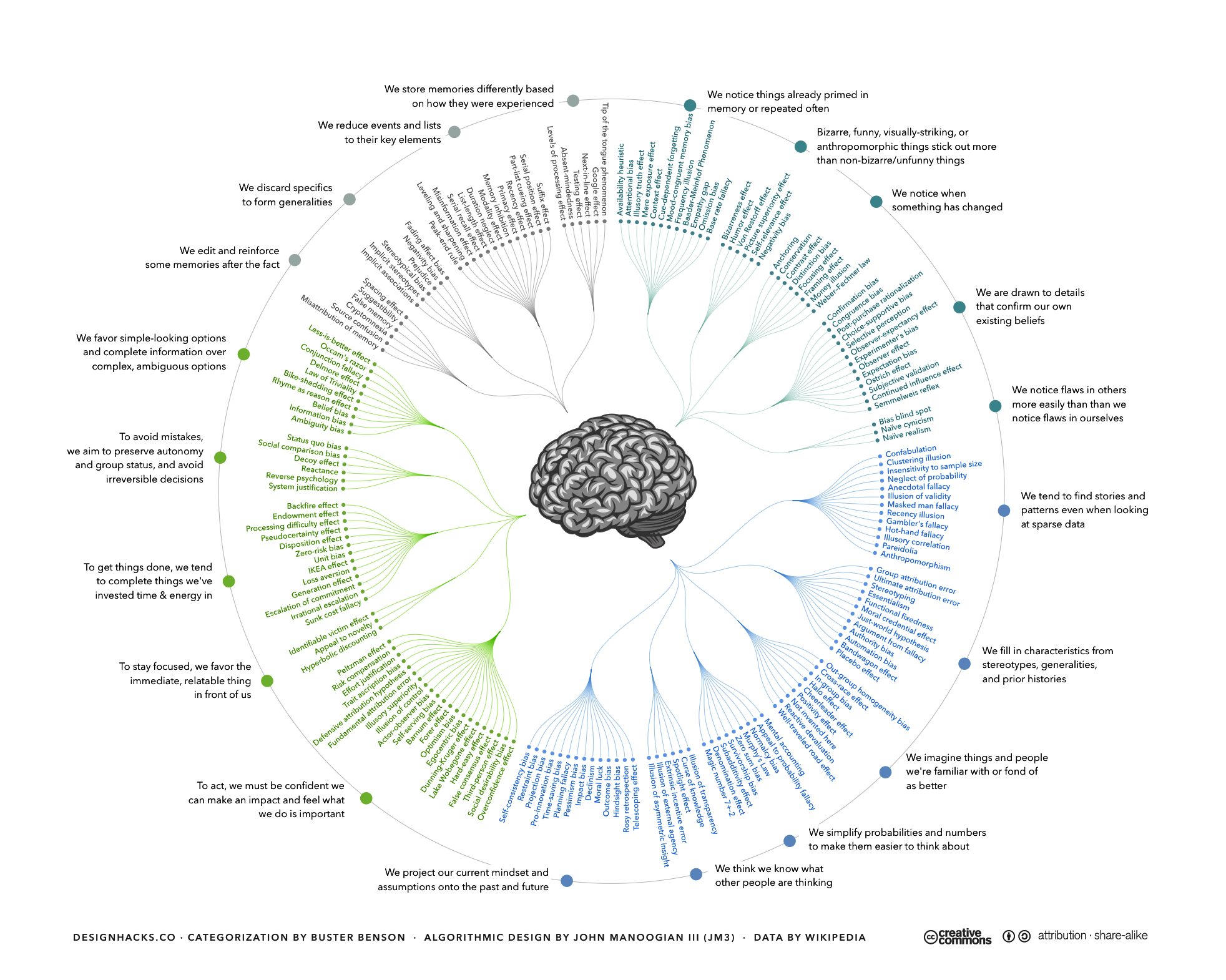}
\caption{\small{Cognitive fallacies listed on Wikipedia. Besides qualitatively categorizing them into classes depending on the context in which they occur, to date there exists no principled way of bringing order to these fallacies, allowing for formally characterizing how one fallacy relates to another.}}
\label{fig_cf_zoo_wiki_map}
\end{figure*}

In the past few decades, empirical investigations into human reasoning led to discovery of new cognitive fallacies, giving rise to a large, ever-growing number of documented fallacies, a state of affairs which can fairly be characterized as a zoo\footnote{The term `cognitive fallacy zoo' is inspired by an analogous terminology in the computational complexity literature, called `complexity zoo.' For details, visit: \texttt{https://complexityzoo.uwaterloo.ca/Complexity\_Zoo}} of cognitive fallacies \citep[e.g.,][]{tversky1973availability,tversky1981framing}. A glance at over a hundred cognitive fallacies listed on Wikipedia (see Fig.~\ref{fig_cf_zoo_wiki_map}) attests to this claim.

In this largely methodological work, we formally present a principled way to bring order to the cognitive fallacy zoo, allowing for a precise characterization of how various fallacies relate to one another. We introduce the idea of establishing an \emph{implication relationship} (IR) (denoted by $\rightsquigarrow$) between a pair of cognitive fallacies, formally characterizing how the occurrence of one fallacy implies another. More formally, for two cognitive fallacies $A,B$, the expression $A \rightsquigarrow B$ denotes that $A$ leads to $B$, i.e., the occurrence of $A$ logically implies the occurrence of $B$. As a proof-of-concept, we present several examples of IRs involving experimentally well-documented cognitive fallacies:~base-rate neglect \citep{tversky1981evidential}, availability bias (Kahneman \& Tversky, 1973), conjunction fallacy (Kahneman \& Tversky, 1983), decoy effect (Huber, Joel, \& Puto, 1982), framing effect \citep{tversky1981framing} and Allais paradox (Allais, 1953).

The idea of establishing IRs between pairs of cognitive fallacies is analogous to, and partly inspired by, the foundational concept of {reduction} in computational complexity theory (see Karp, 1972; Papadimitriou, 2003; Arora \& Barak, 2009; Sipser 2006), which has played a profound role in theoretical computer science, {allowing us to formally establish how various computational problems relate to each other and how the solution to one sheds light on that of another.} After a brief discussion on the role of reduction in computational complexity, we return to the formal characterization of the notion of IR and subsequently present several examples of IRs involving experimentally well-documented cognitive fallacies. But first, some historical notes on reduction in computational complexity.

\section{Reduction in Computational Complexity}
The notion of reduction plays a fundamental role in computational complexity theory, and in theoretical computer science more generally. Informally put, a computational problem $A$ is \emph{reducible} to  computational problem $B$, if every instance of $A$ can be transformed into an instance of $B$. Therefore, the reduction of $A$ to $B$ offers an indirect way of solving $A$, by first reducing $A$ to $B$, and then solving $B$. 

To further clarify the idea of reduction, we provide two examples. As a first example, consider two well-known computational problems, namely, \textsc{Hamiltonian-Path} and \textsc{Hamiltonian-cycle}. The \textsc{Hamiltonian-Path} problem is defined as follows: given a (directed) graph $G$, is there a path which visits every node of $G$ exactly once? The \textsc{Hamiltonian-cycle} is defined as follows: given a (directed) graph $G$, is there a cycle which visits every node of $G$ exactly once? It turns out that \textsc{Hamiltonian-Cycle} is reducible to \textsc{Hamiltonian-Path}. Given that the definitions \textsc{Hamiltonian-Cycle} and \textsc{Hamiltonian-Path} are closely related (since a cycle is a path with its endpoints coinciding), this reduction is not especially surprising. 

As a second example, let us consider \textsc{Hamiltonian-Path} together with the \textsc{3-Colorability} problem, defined as follows: given a graph $G$ and $3$ distinct colors, can you color the nodes of $G$ such that the endpoints of every edge are colored differently? At fist glance, the \textsc{Hamiltonian-Path} and \textsc{3-Colorability} appear to have no connection with one another whatsoever. Surprisingly, however, it turns out that \textsc{Hamiltonian-Path} can be reduced to \textsc{3-Colorability}.\footnote{The reduction can be established by a chain of straightforward reductions: \textsc{Hamiltonian-Path} to \textsc{SAT}, \textsc{SAT} to \textsc{3SAT}, and finally, \textsc{3SAT} to \textsc{3-Coloring}.} Thus, the question of whether a graph $G$ has a Hamiltonian path can be resolved by answering if a corresponding graph $G'$ is 3-colorable. 

\begin{figure*}[h!]
    \centering
    \subfloat[]{{\includegraphics[trim = 0mm 0mm 0mm 0mm,clip,width=0.6\textwidth]{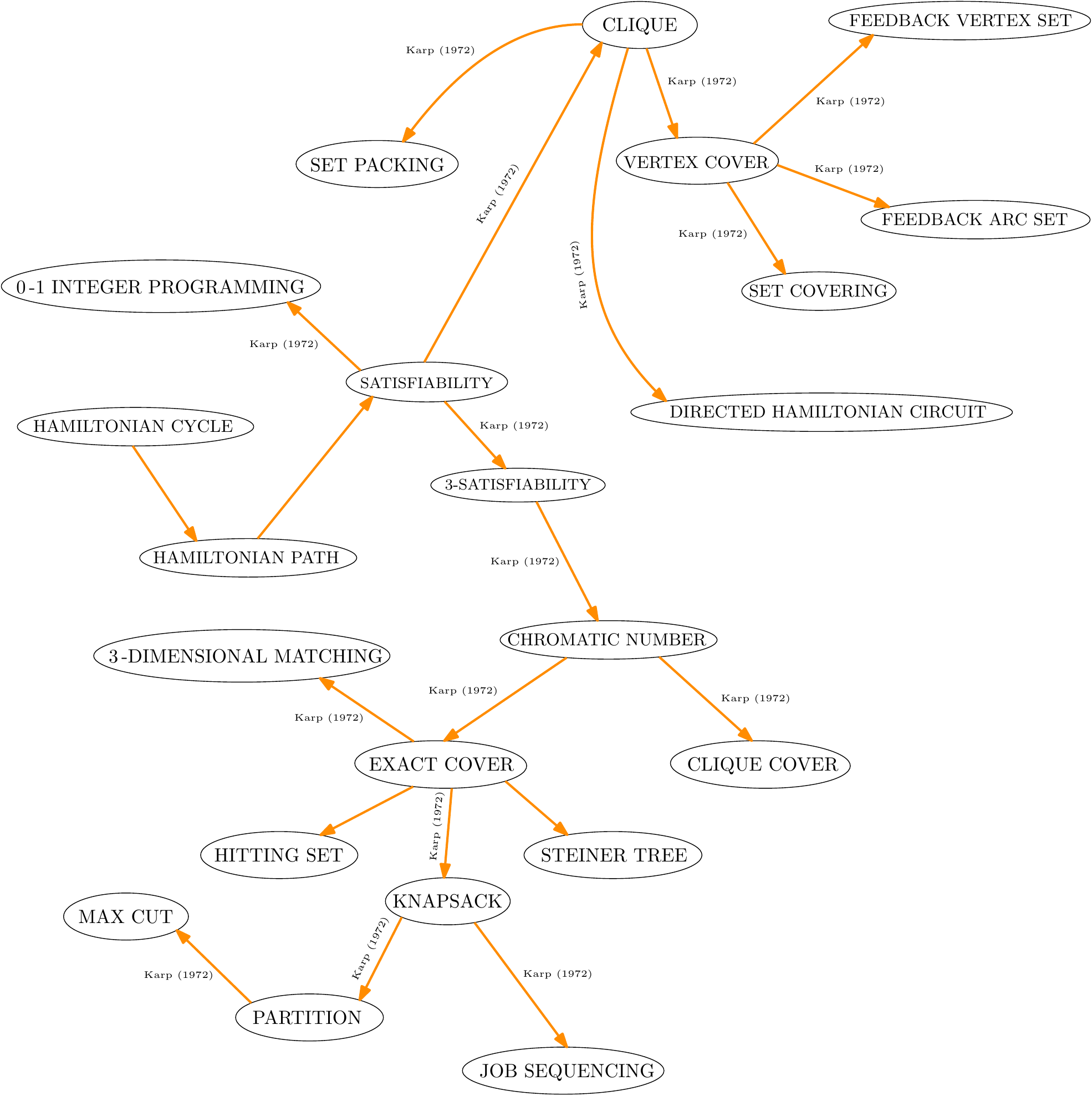} }}
    \hspace*{1pt}
    \subfloat[]{{\includegraphics[trim = 0mm -30mm 0mm 0mm,clip,width=0.37\textwidth]{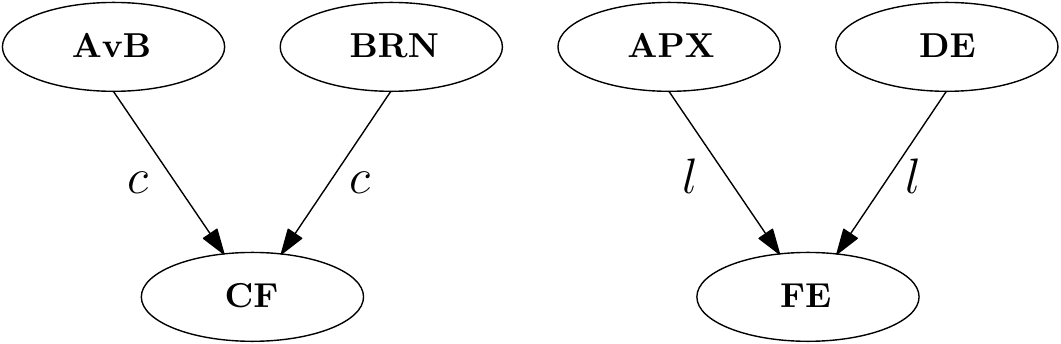} }}
    \hspace*{1pt}
    \linebreak
    \caption{\small{\textbf{(a)} A map showing reductions (directed gold lines) between a set of important computational problems (ovals) in theoretical computer science, formally characterizing how one problem is related to another. The first demonstration of a particular reduction from one problem to another is listed on the corresponding arrow between the two. \textbf{(b)} The IRs formally established in this paper, among well-known cognitive fallacies (AvB: Availability bias, BRN: base-rate neglect, CF: conjunction fallacy, APX: Allais paradox, DE: Decoy effect, FE: framing effect). Letters $c$ (for causal) and $l$ (for logical) on the arrows specify the type of an IR; see the Discussion section for details.}}
    \label{fig_tcs_reduction_np}
\end{figure*}

The idea of reduction has had profound implications for theoretical computer science, allowing for formally connecting seemingly unrelated computational problems to one another (see Fig.~\ref{fig_tcs_reduction_np}(a)). Had reduction not been introduced into theoretical computer science, every computational problem would have had to be investigated on its own, because the solution to one would not have shed any light on the solution to others. It was a major breakthrough when Richard Karp (1972) showed that a key computational problem called \textsc{Satisfiability} could be reduced to a number of other well-known computational problems, a contribution for which he was eventually awarded the Turing award in 1985. It is also worth noting that the (in)famous $\mathcal{P}$~vs.~$\mathcal{NP}$ problem in theoretical computer science, at its core, concerns the possibility or impossibility of establishing a particular form of reduction.

One might wonder if an idea broadly analogous to reduction in computational complexity could be introduced into cognitive science, allowing for formally connecting seemingly different cognitive fallacies with one another, and, hence, bring order to the cognitive fallacy zoo in a principled manner. Primarily motivated by this, and, by analogy with the notion of reduction in theoretical computer science, we introduce the idea of establishing IRs between cognitive fallacies, formally characterizing how one fallacy would imply another.

\section{Implication Relationships: Formalization}
In what follows, we first formally introduce the idea of establishing an \emph{implication relationship} (IR) (denoted by $\rightsquigarrow$) between a pair of cognitive fallacies, followed by several examples of IRs involving experimentally well-documented cognitive fallacies. 

\textbf{Definition (Implication Relationship).} For two cognitive fallacies/biases $A,B$, the fallacy $A$ is said to \emph{implicate} the fallacy $B$ (denoted by $A \rightsquigarrow B$) if and only if the occurrence of $A$ \emph{logically implies} the occurrence of $B$.

\section{Examples on Implication Relationships}
As a proof-of-concept, We next present several examples of IRs involving experimentally well-documented cognitive fallacies, namely, base-rate neglect \citep{tversky1981evidential}, availability bias (Kahneman \& Tversky, 1973), conjunction fallacy (Kahneman \& Tversky, 1983), decoy effect (Huber et al., 1982), framing effect \citep{tversky1981framing}, and Allais paradox (Allais, 1953).

\subsection{Case Study~1: Decoy Effect $\rightsquigarrow$ Framing Effect}
\label{sec_de_to_fe}
As our first example, we formally establish an IR between two well-documented cognitive fallacies, namely, the decoy effect (DE) and the framing effect (FE).

\textbf{The Framing Effect (FE):} If people produce different responses for two equivalent tasks, the framing effect (FE) has occurred \citep{tversky1981framing,kahneman1984choices}. In that light, FE is a violation of the extensionality principle (Bourgeois-Gironde \& Giraud, 2009), which prescribes that two equivalent tasks should elicit the same response.

FE is well captured by \citet{tversky1981framing}: Subjects were asked to ``imagine that the U.S. is preparing for the outbreak of an unusual Asian disease, which is expected to kill $600$ people. Two alternative programs to combat the disease have been proposed. Assume the exact scientific estimate of the consequences of the programs are as follows." In one condition, subjects were presented with a choice between Programs A and B, while in another condition, subjects were asked to choose between Programs C and D:
\begin{itemize}
\item[] \textbf{Program A:} $200$ people will be saved.\vspace*{-5pt}
\item[] \textbf{Program B:} There is a $\frac{1}{3}$ probability that 600 people will be saved, and a $\frac{2}{3}$ probability that no people will be saved.\vspace*{-5pt}
\item[] \textbf{Program C:} $400$ people will die. \vspace*{-5pt}
\item[] \textbf{Program D:} there is a $\frac{1}{3}$ probability that nobody will die, and a $\frac{2}{3}$ probability that 600 people will die.
\end{itemize}
Despite the equivalence of these Programs pairs, a majority of the first group preferred Program A ($=\text{C}$), while a majority of the second group preferring Program D ($=$ B).  

\textbf{The Decoy Effect (DE):} The decoy effect (DE) refers to a change in people's preference between two options, when presented with a third \emph{asymmetrically-dominated} option, i.e., an option which is inferior to one option in all respects, but, in comparison to the other option, it is inferior in some respects and superior in others. In that light, DE is a violation of the independence of irrelevant alternative axiom of rational decision theory (Ray, 1973), which prescribes the following: If $A$ is preferred to $B$ out of the choice set $\{A,B\}$, introducing a third option $X$, hence expanding the choice set to $\{A,B,X\}$, should not make $B$ preferable to $A$. 

We are now well-positioned to formally present our result. 

\textbf{Proposition~1.} \emph{The following holds:}
\begin{eqnarray*}
\text{DE} \rightsquigarrow \text{FE}.
\end{eqnarray*}

\emph{\textbf{Proof.}}  According to normative principles, preference for the choice sets $\{A,B\}$ and $\{A,B,X\}$ should be the same, with $X$ being an asymmetrically-dominated option. The rationale is the following: Since $X$ is inferior to one option in \textit{all} respects, rationally $X$ should never be chosen; hence, the preference pattern for the choice sets $\{A,B\}$ and $\{A,B,X\}$ should be identical. Therefore, whenever people's preference pattern for the choice sets $\{A,B\}$ and $\{A,B,X\}$ differs (which is the case for DE), it logically implies the violation of the extensionality principle, hence granting the occurrence of FE. This concludes the proof. \hfill $\blacksquare$

The message of Proposition~1 is simple: From the standpoint of normative principles, the two choice sets $\{A,B\}$ and $\{A,B,X\}$ (with $X$ being an asymmetrically-dominated option) are equivalent, therefore people's showing different preference patterns for the two choice sets, as is the case in DE, is a clear indication of FE. Proposition~1, therefore, formally establishes that the occurrence of DE leads to the occurrence of FE, that is, whenever DE occurs, so does FE.

\subsection{Case Study~2: Base-Rate Neglect $\rightsquigarrow$ Conjunction Fallacy}
\label{sec_brn_to_cf}
As our second example, we formally establish an IR between another pair of well-documented cognitive fallacies, namely, the base-rate neglect (BRN) and the conjunction fallacy (CF). BRN and CF can be characterized as follows.

\textbf{The Base-Rate Neglect (BRN):} Base-rate neglect (BRN) \citep{tversky1981evidential} refers to people not considering prior probabilities in their judgments under uncertainty.

\textbf{The Conjunction Fallacy (CF):} For two events $A,B$ and presented with evidence $e$, people judge the probability of the event $A\cap B$ to be greater than that of $A$ (or $B$), in isolation. That is, more formally, people judge: $\bP(A\cap B|e)>\bP(A|e)$. In that light, CF is a clear violation of the axioms of probability (since $\forall A,B,\ A\cap B \subseteq A \Rightarrow \bP(A\cap B|e)\leq \bP(A|e)\ \forall e\neq \varnothing$; that is, the probability of a subset of $Y$, in principle, cannot be greater than that of $Y$). 

CF is well captured in the famous Linda experiment by Tversky and Kahneman~(1981). Presented with a description ($e$) of Linda, a politically active, single, outspoken, and very bright 31-year-old female, people overwhelmingly judge that Linda is more likely to be a feminist bankteller ($A\cap B$) than to be a bankteller ($A$).

We are now well-positioned to formally present our result. 

\textbf{Proposition~2.} \emph{The following holds.}
\begin{eqnarray*}
\text{BRN} \rightsquigarrow \text{CF}.
\end{eqnarray*}

\textbf{\emph{Proof.}} Since $\bP(A\cap B|e)=\bP(e|A\cap B)\bP(A\cap B)$ and $\bP(A|e)=\bP(e|A)\bP(A)$, we have:
\begin{eqnarray*}
\dfrac{\bP(A\cap B|e)}{\bP(A|e)}=\dfrac{\bP(e|A\cap B)}{\bP(e|A)} {\dfrac{\bP(A\cap B)}{\bP(A)}},
\end{eqnarray*}
where the term $\frac{\bP(A\cap B)}{\bP(A)}$ indicates the ratio between priors $\bP(A\cap B)$ and $\bP(A)$. If BRN occurs (which results in the term $\frac{\bP(A\cap B)}{\bP(A)}$ being dropped), it follows that:
\begin{eqnarray*}
\dfrac{\bP(A\cap B|e)}{\bP(A|e)}=\dfrac{\bP(e|A\cap B)}{\bP(e|A)}.
\end{eqnarray*}

Assuming that $\bP(e|A\cap B)>\bP(e|A)$, which is the case in the context of CF (see the Linda experiment discussed above), it follows that:
\begin{eqnarray*}
\dfrac{\bP(A\cap B|e)}{\bP(A|e)}=\dfrac{\bP(e|A\cap B)}{\bP(e|A)}>1 \Rightarrow \bP(A\cap B|e) > \bP(A|e),
\end{eqnarray*}
hence CF occurs. This completes the proof. \hfill $\blacksquare$

In simple terms, Proposition~2 shows that the occurrence of BRN leads to the occurrence of CF, i.e., BRN gives rise to CF.

\subsection{Case Study~3: Allais Paradox $\rightsquigarrow$ Framing Effect}
\label{sec_apx_to_fe}
As our third example, we formally establish an IR between the Allais paradox (APX) and FE. APX can be characterized as follows. (See Sec.~\ref{sec_de_to_fe} for the characterization of FE.)

\textbf{The Allais Paradox (APX):} The Allais paradox refers to an observed reversal in participants' choices in two different experiments, each of which consists of a choice between two gambles, $A$ and $B$, while in fact, according to the independence axiom of rational decision-making (Von Neumann \& Morgenstern, 1953), no such a reversal should occur. That is, although the independence axiom grants the equivalence of the two experiments, the pattern of people's preference nevertheless reverses from the first experiment to the second.\footnote{The reader is referred to Allais~(1953) for a clear description of the two experiments.}

\textbf{Proposition~3.} \emph{The following holds:}
\begin{eqnarray*}
\text{APX} \rightsquigarrow \text{FE}.
\end{eqnarray*}

\textbf{\textit{Proof.}} The proof is evident from the characterization of APX given above. Although the independence axiom of rational decision-making (Von Neumann \& Morgenstern, 1953) grants the equivalence of the two experiments entertained in APX, the pattern of people's preference nevertheless reverses from one to the other. That is, in the case of APX, people produce different responses for two equivalent experiments. Therefore, the occurrence of the Allais paradox logically implies the occurrence of the framing effect. This concludes the proof. \hfill $\blacksquare$ 

\subsection{Case Study~4: Availability Bias $\rightsquigarrow$ Conjunction Fallacy}
\label{sec_avb_to_cf}
As our final example, we formally establish an IR between the well-documented Availability bias (AvB) and CF. AvB can be concisely characterized as follows: (See Sec.~\ref{sec_brn_to_cf} for the characterization of CF.)

\textbf{The Availability Bias (AvB):} Extreme events come to mind easily, people overestimate their probabilities, and over-represent them in decision-making \citep{tversky1973availability,lieder2018overrepresentation,Nobandegani2018}. Formally, people overestimate the probability of an event $o$, $p(o)$, proportional to the absolute value of its subjective utility $u(o)$ (Lieder et al., 2018; Bordalo, Gennaioli, \& Shleifer, 2012). That is, people's subjective probability of event $o$, $q(o)$, is given by\footnote{We must emphasize that our establishing of the IR between AvB and CF only depends on the broad assumption that the more extreme an event is, the more people overestimate its probability, and holds for any $q(o)$ which satisfies this condition, e.g., $q(o)\propto p(o)|u(o)|\sqrt{\frac{1+|u(o)|\sqrt{s}}{|u(o)|\sqrt{s}}}$ \citep{Nobandegani2018}. Therefore, the assumption $q(o)\propto p(o)|u(o)|$ made in the characterization of AvB is only one choice out of infinitely-many possibilities satisfying the said condition, and hence, is not necessary.} $q(o)\propto p(o)|u(o)|$. 

\textbf{Proposition~4.} \emph{Let $o_1$ and $o_2$ be two events, and let $o_\wedge$ denote the event corresponding to the occurrence of $o_1$ and $o_2$ together, i.e., the one corresponding to the conjunction of the two events $o_1$ and $o_2$. Assuming that $\forall i=1,2,\ |u(o_\wedge)|\gg |u(o_i)|$, the following holds:}
\begin{eqnarray*}
\text{AvB} \rightsquigarrow \text{CF}.
\end{eqnarray*}

\textbf{\emph{Proof.}} According to the characterization of AvB given above, $q(o_\wedge)\propto p(o_\wedge)|u(o_\wedge)|$ and $\forall i=1,2,\ q(o_i)\propto p(o_i)|u(o_i)|$. We have,
\begin{equation*} 
\forall i=1,2,\quad \dfrac{q(o_\wedge)}{q(o_i)}=\dfrac{p(o_\wedge)}{p(o_i)} \dfrac{|u(o_\wedge)|}{|u(o_i)|}.
\end{equation*}
It follows from the axioms of probability that $\forall i=1,2,\ p(o_\wedge)\leq p(o_i)$; hence, $\forall i=1,2,\ \frac{p(o_\wedge)}{p(o_i)}\leq 1$. However, since $\forall i=1,2,\ |u(o_\wedge)|\gg |u(o_i)|$, it follows that $\frac{|u(o_\wedge)|}{|u(o_i)|}\gg 1,\ \forall i=1,2$. Therefore, altogether, $\forall i=1,2,\ \frac{q(o_\wedge)}{q(o_i)}>1$ which implies $\forall i=1,2,\ {q(o_\wedge)}>{q(o_i)}$, granting the validity of the conjunction fallacy (CF). This concludes the proof. \hfill $\blacksquare$

The message of Proposition~4 is simple. If people judge the conjunction of two events to be much more extreme than each of them individually (i.e., $\forall i=1,2,\ |u(o_\wedge)|\gg |u(o_i)|$), then the occurrence of AvB leads to the occurrence of CF; that is, whenever AvB happens, so does CF.

\section{General Discussion}
In this largely methodological work, we introduce the notion of implication relation (IR) between a pair of cognitive fallacies, formally characterizing how one would logically imply the other.

A closer examination of Propositions 1 to 4 and their proofs reveals that IRs can be categorized into two broad types: \emph{logical-IRs} (denoted by $\lir$) and \emph{causal-IRs} (denoted by $\cir$). Establishing a logical-IR, $\lir$, from a fallacy $F_1$ to another fallacy $F_2$ implies that $F_1$ is a special case of $F_2$, with every instance of $F_1$ being an instance of $F_2$. For example, a closer examination of Proposition~1 and its proof reveals that DE is a special case of FE, with every instance of DE being an instance of FE in disguise. The same understanding holds for Proposition~3 and its proof, indicating that APX is simply a special case of FE, with every instance of APX being an instance of FE in disguise. Hence, using our newly introduced notation: $\text{DE}\lir\text{FE}$ and $\text{APX}\lir\text{FE}$. Establishing a causal-IR, $\cir$, from a fallacy $F_1$ to another fallacy $F_2$ implies that the occurrence of $F_1$ brings about (i.e., causes) the occurrence of $F_2$. For example, a closer examination of Proposition~2 and its proof reveals that the occurrence of BRN brings about the occurrence of CF, i.e., there is a cause-effect relationship between BRN and CF, with BRN being the cause and CF the effect. The same understanding holds for Proposition~4 and its proof, indicating that the occurrence of AvB brings about the occurrence of CF, i.e., there is a cause-effect relationship between AvB and CF, with AvB being the cause and CF the effect. Hence, using our newly introduced notation: $\text{BRN}\cir\text{CF}$ and $\text{AvB}\cir\text{CF}$. Drawing further on the analogy between IR and reduction in computational complexity, it is worth noting that there also exist several types of reduction in computational complexity, namely, Karp's reduction, Cook's reduction, truth-table reduction, L-reduction, A-reduction, P-reduction, E-reduction, AP-reduction, PTAS-reduction, etc.

Importantly, logical-IRs and causal-IRs have quite different implications. If $F_1\lir F_2$ holds (implying that $F_1$ is a special case of $F_2$ as discussed above), it then follows that a \emph{complete} account of $F_2$ should also account for $F_1$, and, in that sense, accounting for $F_2$ is more demanding\footnote{Accounting for $F_2$ is ``more demanding" than for $F_1$, as a complete account of $F_2$ would necessarily have to explain a wider range of cases, including all instances of $F_1$ as a subset.} than accounting for $F_1$. For example, since DE is a spacial case of FE (see Proposition~1 and its proof), that is, DE is nothing but FE in disguise, any complete account for FE inevitably should also account for DE, implying that accounting for FE is more demanding than accounting solely for a special case of FE, DE. However, if $F_1\cir F_2$ holds (implying that the occurrence of $F_1$ brings about $F_2$), it then follows that an account of $F_1$ naturally serves as an account of $F_2$ due to the following rationale: If $X$ causes $F_1$, and $F_1$ causes $F_2$, it then follows that $X$ causes $F_2$, with $F_1$ serving as a mediator. In that light, establishing causal-IRs between various cognitive biases/fallacies has an intriguing implication: For any chain of causal-IRs $F_1\cir F_2\cir F_3\cir\ldots \cir F_{n-1}\cir F_n$, any mechanistic account of $F_i$ naturally serves as an account of $F_{i+1}, F_{i+2},\ldots, F_n$. For example, since the occurrence of BRN causes the occurrence of CF (see Proposition~2 and its proof), it then follows that any mechanistic account of BRN naturally serves as an account of CF, with BRN serving as a mediator. This understanding has a very intriguing implications for studies of cognitive fallacies in general: Establishing a chain of causal-IRs $F_1\cir F_2\cir F_3\cir\ldots \cir F_{n-1}\cir F_n$, clearly reveals which of the fallacies $F_1,\cdots,F_n$ is more pivotal or fundamental to account for; the answer is of course the left-most fallacy in the chain, i.e., $F_1$. This strongly suggests that, directing efforts toward finding a comprehensive, satisfying account of $F_1$ would be the most rewarding research agenda, because, thanks to the established chain of causal-IRs, we would get a set of comprehensive, satisfying accounts of all $F_2, F_3, \cdots, F_n$ for free! Therefore, identifying IRs could {systematize} and {guide} a research agenda, with a huge increase in research efficiency.

Proposition~4, establishing $\text{AvB} \rightsquigarrow \text{CF}$, demonstrates an interesting possibility wherein, under a set of auxiliary assumptions (e.g.~$\forall i=1,2,\ |u(o_\wedge)|\gg |u(o_i)|$ in this case), an IR can be established between two fallacies. The idea of establishing IRs under a set of assumptions widens the applicability of the notion of IR, allowing it to link together pairs of cognitive fallacies that would have little connections unless further assumptions are invoked. Drawing again on the analogy between IR and reduction in computational complexity, it is worth noting that in establishing reductions it is common practice to evoke various assumptions/constraints on the characterization of computational problems (e.g.~$3$-\textsc{Sat}\linebreak instead of \textsc{Sat}) and/or on the forms of reductions themselves (e.g.~\emph{polynomial-time} reductions or \emph{linear-time} reductions). Importantly, these auxiliary assumptions should be empirically confirmed, motivating new and exciting experimental avenues of research. Empirical confirmations of such auxiliary assumptions, empirically justifies the validity of invoking such assumptions. Importantly, empirical disconfirmation of such assumptions, of course, discredit the said established IR, inviting attempts for establishing other IRs (in the hope that they would survive empirical tests), or for invoking other empirically validated assumptions which would save the established IR, motivating new theoretical and empirical work.

In this work, as a proof-of-concept, we established IRs between several well-documented cognitive biases; see Fig.\ref{fig_tcs_reduction_np}(b). Future work should investigate the possibility of establishing IRs between a wider range cognitive biases/fallacies, with the ultimate goal of developing a principled, comprehensive map of cognitive biases/fallacies, broadly resembling what is shown in Fig.~\ref{fig_tcs_reduction_np}(a) in the context of computational complexity. While many questions remain open, and much work is left to be done in this direction, we hope to have made some progress toward systematically bringing order to the cognitive fallacy zoo. We see our work as a first step in this direction. 

Much like the fundamental concept of reduction in computational complexity, as eminently featured in the work by Karp~(1972), that revolutionized the fields of computational complexity and, arguably, theoretical computer science as a whole, we hope the pursuit of the vision outlined in our paper to have a comparable impact on the fields of cognitive science and cognitive psychology.

\vspace*{5pt}
\hspace*{-10pt}\textbf{Acknowledgments}~\small{We would like to thank Marcel~Montrey, Kevin~da~Silva~Castanheira, and Peter Helfer for helpful comments on an earlier draft of this work. This work is supported by an operating grant to TRS from the Natural Sciences and Engineering Research Council of Canada.}

\nocite{simon1972theories}
\nocite{tversky1974judgment}
\nocite{papadimitriou2003computational}
\nocite{arora2009computational}
\nocite{sipser2006introduction}
\nocite{allais1953}
\nocite{bourgeois2009framing,kahneman1984choices,tversky1981framing,tversky1981evidential,tversky1973availability}
\nocite{tversky1983extensional}
\nocite{huber1982adding,karp1972reducibility,ray1973independence}
\nocite{kahneman2011thinking}
\renewcommand\bibliographytypesize{\tiny}
\bibliographystyle{apacite}
\bibliography{ref}
\end{document}